\documentclass[11pt,a4paper]{article}
\pdfoutput=1   
\usepackage[utf8]{inputenc}
\usepackage{tabularx}
\usepackage{mathtools}
\usepackage{xspace}
\usepackage{slashed}
\usepackage{alpha} 
\hypersetup{%
   pdftitle    = {Symanzik Improvement with Dynamical Charm: A 3+1 Scheme for Wilson Quarks},
   pdfauthor   = {Patrick Fritzsch, Rainer Sommer, Felix Stollenwerk, Ulli Wolff},
   pdfkeywords = {Lattice QCD, Schr\"odinger Functional, O(a) improvement}
}%
\usepackage{defs}
\usepackage{bbm}
\begin{document}

\preprintno{%
CERN-TH-2018-084\\
DESY~18-058\\
HU-EP-18/10\\[5em]
}

\title{%
Symanzik Improvement with Dynamical Charm:\\ A 3+1 Scheme for Wilson Quarks 
}

\collaboration{\includegraphics[width=2.8cm]{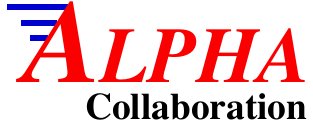}}

\author[cern]{Patrick Fritzsch}
\author[desy,hu]{Rainer~Sommer}
\author[hu]{Felix Stollenwerk}
\author[hu]{Ulli Wolff}

\address[cern]{CERN, Theoretical Physics Department, 1211 Geneva 23, Switzerland}
\address[desy]{John von Neumann Institute for Computing (NIC), DESY, Platanenallee~6, 15738 Zeuthen, Germany}
\address[hu]{Institut~f\"ur~Physik, Humboldt-Universit\"at~zu~Berlin, Newtonstr.~15, 12489~Berlin, Germany}

\begin{abstract}
We discuss the problem of lattice artefacts in QCD simulations enhanced by the
introduction of dynamical charmed quarks. In particular, we advocate the use of
a massive renormalization scheme with a close to realistic charm mass. To
maintain O($a$) improvement for Wilson type fermions in this case we define a
finite size scheme and carry out a nonperturbative estimation of the clover
coefficient $\csw$. It is summarized in a fit formula $\csw(g_0^2)$ that
defines an improved action suitable for future dynamical charm simulations.
\end{abstract}


\maketitle

\section{Introduction}\label{sec:intro}

Physicists have been rather fortunate that perturbation theory in the form of
low order Feynman diagrams has been largely sufficient to practically establish
today's highly successful Standard Model of elementary particles from
experiment.  Due to asymptotic freedom even the strongly interacting sector of
confined quarks and gluons (QCD) could be pinned down in this way, based on
high energy scattering of weakly interacting probes (e.g. $e^+\,e^-$ and $e^-
p$). The thus completely characterized theory is expected to also produce the
hadronic bound states observed in nature. To extract these predictions---and
thus ultimately validate the complete theory---nonperturbative techniques
become indispensable.

The only known nonperturbative definition of QCD is the regularization on a
lattice which in particular provides a systematic way to extract
nonperturbative information about hadrons from the theory by Monte Carlo
simulations of QCD on sequences of finite computational lattices. This inflicts
however several unavoidable distortions of the theory which need to be
controlled.  The lattice spacing $a$ acts as an ultraviolet regulator that has
to be extrapolated to zero up to tolerable errors. A finite system length $L$
is introduced and has to be made effectively infinite in a similar sense.
Finally---as in experiment---all predictions come with statistical errors that
have to be small enough for significant comparisons.  In particular the
symmetries of QCD  get modified by the lattice regularization and some of them
only re-emerge in the continuum limit $a\to 0$. Very obviously this holds true
for infinitesimal translation invariance. Wilson~\cite{Wilson:1974sk} showed
that gauge invariance can be preserved on the lattice. It is deemed to be so
essential for physics that only such formulations are considered.  For the
gluons this is achieved by writing the action in terms of parallel transporters
around plaquettes or other small closed loops as we shall detail below.

For the quark degrees of freedom more subtle questions arise. The naive gauge
covariant discretization of the first derivative in the Dirac action leads to
fermion doubling, i.e., an unwanted proliferation of the degrees of freedom. It
has turned out that this is in fact related to a certain principal
incompatibility between the lattice discretization and chiral
invariance~\cite{Nielsen:1981hk}, which is a very important approximate
symmetry in QCD. There are several possibilities to coexist with the
corresponding no-go theorem~\cite{Nielsen:1981hk}.  The present article refers
to the choice proposed also by Wilson~\cite{Wilson:1975id} which nowadays is
called Wilson fermions. By adding the so-called Wilson term to the action (see
below) the one-to-one correspondence between lattice Dirac fields and physical
quarks is restored and the vectorial flavor symmetries are intact even in the
presence of the lattice.  There is however a price to pay for this bonus. The
Wilson term violates chiral symmetry, which must emerge non-trivially in the
continuum limit. Moreover, without further precautions, the lattice cutoff
effects vanish only asymptotically linearly in $a$. The gluon action alone, and
also certain other fermion discretizations, lead to a behavior proportional to
$a^2$ which much facilitates the extrapolation and enhances the precision in
and the reliability of this step. A further complication due to the missing
chiral invariance at finite $a$ is the need to control an additive quark mass
renormalization. We only briefly mention here that technically the dynamical
quarks dominate the simulation costs. The degree to which this is the case
depends on the chosen discretization with Wilson fermions being one of the
cheaper options.

To limit the simulation costs, lattice calculations have proceeded through a
series of approximations concerning the quantum effects due to virtual
dynamical flavor contributions. At first they were neglected altogether in the
quenched approximation.  Then, to have a unitary theory and to capture the most
important loop effects, two or three light flavors were included.  In this
article we are concerned with the next level where we add the dynamical charmed
quark.  It has a much larger mass $\mcharm$ compared to up, down or strange
quarks, and this can lead to cutoff effects proportional to the dangerously
large dimensionless factor $a\mcharm$. Special measures toward their control
are in the focus of this article. 
In particular we suggest and formulate Symanzik improvement
of the Wilson theory in a massive renormalization scheme. The 
Sheikholeslami-Wohlert coefficient~\cite{Sheikholeslami:1985ij} 
is determined non-perturbatively with a massive charm quark.

Massive renormalization schemes are not new in the context of 
a reduction of $\rmO(\!(am_\charm)^n\!)$ effects. They have been used 
(sometimes implicitly) for space-time symmetric formulations \cite{Baron:2010bv,Carrasco:2014cwa,Follana:2006rc,Hart:2008sq,Bazavov:2010ru}
and with broken space-time symmetry 
\cite{ElKhadra:1996mp,Aoki:2001ra,Harada:2001ei,Harada:2001fi,Harada:2001fj,Harada:2002rf,Aoki:2003dg,Aoki:2004th,Christ:2006us,Lin:2006ur,Oktay:2008ex}. The latter arise from
considerations of effective field theories for heavy quarks and
have so far been applied to quenched quarks or heavy quarks treated 
in a mixed action approach. References \cite{ElKhadra:1996mp,Aoki:2001ra,Harada:2001ei,Harada:2001fi,Harada:2001fj,Harada:2002rf,Aoki:2003dg,Aoki:2004th,Christ:2006us,Lin:2006ur,Oktay:2008ex}
aim at a reduction of the quadratic and higher, $n\geq 2$, terms
by tree-level and one-loop perturbation theory. We here focus
just on the non-perturbative removal of the linear term in $am_\charm$.

This paper is a highly condensed summary of the PhD
thesis~\cite{Thesis:Stollenwerk} by Felix Stollenwerk. In sect.~2 we review
on-shell Symanzik O($a$) improvement followed by the comparisons of mass
dependent and independent renormalization schemes in sect.~3. In sect.~4 we
define a line of constant physics by maintaining a set of finite size
observables as $a$ is changed.  In sect.~5 this is employed for a
nonperturbative determination of the improvement coefficient $\csw$ in a scheme
of three light flavors together with the close to physical dynamical charm
quark. A fit formula is presented that allows to implement this action in
future large volume simulations at and close to the physical point.  We end
with some conclusions in sect.~6.

\section{Symanzik improvement}\label{sec:SymanzikImpr}

Symanzik has put forward a
method~\cite{Symanzik:1981hc,Symanzik:1983dc,Symanzik:1983gh} to accelerate the
continuum limit in lattice field theories by increasing the power of $a$
characterizing the leading lattice artifacts. It is routinely applied to Wilson
fermions to achieve an $a^2$ continuum scaling.  This so-called Symanzik
improvement programme starts from an effective {\em continuum} theory which
first only describes a given lattice theory including its leading order
artefacts. In a renormalizable quantum field theory we are at first instructed
to include in the action a combination of all possible terms up to (including)
mass dimension four which possess the desired symmetry. They typically are
small in number and are related to the number of free (renormalized)
parameters.  To reproduce the leading cutoff effects of O($a$), the scope has
to be widened to all such operators of dimension five. They appear multiplied
by dimensionless coefficient functions times an explicit factor of $a$.  They
could be determined by imposing a sufficient number of conditions where
observables of the lattice theory are matched. These are analogous to
renormalization conditions to fix the relevant terms up to dimension four.  In
a next step one then modifies the lattice action such that the dimension five
terms in the associated effective continuum theory become zero.  This can be
achieved by including in the lattice theory a linear combination of discretized
versions of the dimension five operators with suitably tuned coefficients. Any
ambiguity in this step only affects the uncancelled higher order artefacts.
What is exploited here is the structure of the effective theory which allows to
describe all O($a$) artefacts by a finite set of parameters by organizing all
possible terms according to their naive dimension. This could in principle be
invalidated by large anomalous dimensions due to quantum corrections.  For
asymptotically free theories like QCD this does however not happen to any
finite order of perturbation theory. In this sense the Symanzik programme is
expected to work in a similar way as renormalizability and thus the existence
of the continuum limit.

In the following we shall restrict our focus on the simpler on-shell Symanzik
improvement~\cite{Luscher:1996sc,Luscher:1984xn}.  This means that we only
improve observables that are derived from correlation functions with operators
separated by physical distances that remain finite in the continuum limit. Then
the use of the quantum equations of motion is justified which allows to relate
dimension five operators to each other and thus to reduce the number of
independent terms that have to be managed. As we shall report below we then
need, apart from terms related to quark masses, only one single term, the Pauli
or, on the lattice, the clover-leaf term~\cite{Sheikholeslami:1985ij} to cancel
O($a$) artefacts in QCD.

We now summarize the structure of the Symanzik effective (continuum) Lagrangian
that describes lattice QCD with $\Nf$ flavors of Wilson quarks with a diagonal
mass matrix $M$, including O($a$) lattice artefacts,
\begin{align}
  \mathcal{L}_\mathrm{Sym} &=
                -\tfrac{1}{2g_0^2} \,\tr[F_{\mu\nu} F_{\mu\nu}] 
                + \bar{\psi} \slashed{D} \psi
                + \bar{\psi} M \psi 
                + \rho \,\tr[M] \,\bar{\psi} \psi                     \nonumber \\
         &\quad + a \sigma_0 \,\bar{\psi} \sigma_{\mu\nu} F_{\mu\nu} \psi
                + a \sigma_1 \,\bar{\psi} M^2 \psi
                + a \sigma_2 \,\tr[M]     \,\bar{\psi} M \psi
                + a \sigma_3 \,\tr{[M^2]} \,\bar{\psi} \psi            \nonumber \\
         &\quad + a \sigma_4 \,(\tr[M])^2 \,\bar{\psi} \psi
                + a \sigma_5 \,\tr[M]\,\tr[F_{\mu\nu} F_{\mu\nu}] \,. \label{Sym_action}
\end{align}
The first line contains the dimension four terms. The anti-Hermitian algebra
valued field strength is denoted as usual by $F_{\mu\nu}$ and $\slashed{D}$ is
the covariant Dirac operator.  A subtlety here is, that due to the missing
chiral symmetry of the lattice action, the adjustable constant $\rho$ is
required to match the renormalization pattern of the lattice theory which
differs for the flavor singlet and non-singlet components in $M$.  The
remaining lines show the dimension five operators.  We have already implemented
simplifications from using the equations of motion (on-shell improvement) here.
In this way we could drop the terms $a\bar{\psi} D^2 \psi$, $a \bar{\psi}
\slashed{D} M \psi$ and $ a\tr[M]\bar{\psi} \slashed{D} \psi$.

Similarly to the action also currents appearing in correlation functions have
to be nontrivially matched by the Symanzik effective theory. One has to allow
for the mixing of terms of the same and the next higher dimension as far as
they share all symmetries preserved by the lattice theory. We here restrict
ourselves to the off-diagonal axial vector and pseudoscalar currents made of
quarks of flavor $i \not= j$,
\begin{align}
 A_{\mu}^{ij} &= \bar{\psi}_i \gamma_{\mu}\gamma_5 \psi_j \,, \quad
 P^{ij}       = \bar{\psi}_i \gamma_5 \psi_j             \,.
 \label{Amucurrent}
\end{align}
Their mixing pattern is
\begin{align}
 (A_\mathrm{Sym})_{\mu}^{ij} &= \left\{1+\omega_1 \,a\tr[M]+\omega_2 \,a(m_i+m_j)\right\}A_{\mu}^{ij} +\omega_0\, a\partial_{\mu} P^{ij} \,, \\
\shortintertext{and} 
 (P_\mathrm{Sym})^{ij} &= \left\{1+\omega_1' \,a\tr[M]+\omega_2' \,a(m_i+m_j)\right\}P^{ij} \,. 
 \label{Pcurrent}
\end{align}

In continuum QCD chiral symmetry implies the PCAC relation for the renormalized
currents. Formally, i.e., disregarding renormalization for the moment, it reads
\begin{equation}
 \partial_{\mu} A_{\mu}^{ij} = (m_i+m_j) P^{ij} \label{PCAC0}
\end{equation}
and it holds as an operator relation. This means that the above identity may be
inserted in correlators with suitable operators $O$
\begin{equation}
 \langle \partial_{\mu} A_{\mu}^{ij} O \rangle = (m_i+m_j) \langle  P^{ij} O \rangle \,.
\end{equation}
Such relations are typically violated by O($a$) artefacts in the lattice
theory.  They thus may serve as a source of relations which allow to determine
coefficients of O($a$) corrections in the Symanzik effective theory and
ultimately in the improved lattice theory.

\section{Mass (in)dependent renormalization scheme and improvement}\label{sec:schemes}

In mass independent renormalization schemes for QCD the conditions that define
the renormalized theory are formulated at vanishing quark masses in terms of an
external renormalization scale $\mu$ only.  This in particular leads to
relatively simple renormalization group equations with $\beta$- and
$\gamma$-functions depending on the dimensionless coupling only.  In the
technically convenient finite size schemes the system size $L$ is often used as
external scale. In this article we shall argue that in the presence of
dynamical charmed quarks, mass dependent schemes, where renormalization factors
are allowed to depend on quark masses, offer practical advantages.

\subsection{Mass independent case}

In the case of mass independent renormalization schemes the relation between
bare and renormalized gauge coupling reads (without improvement)
\begin{equation}
 \gR^2 = Z_g(g_0^2,a\mu) g_0^2, \label{grm0}
\end{equation}
and $Z_g$ is determined in this way, once a suitable $\gR$ has been defined.
The relation between bare and renormalized masses has the structure
\begin{equation}
 \mR[,i] = Z_m(g_0^2,a\mu)\left[ \mqi + (r_m(g_0^2)-1)\,\tr[\Mq]/\Nf \right] \label{mrm0}
\end{equation}
involving the bare subtracted quark masses of flavor $i$
\begin{equation}
 \mqi = m_i - \mcr(g_0^2)\,, \quad \Mq=\diag(m_{\rm q,1},\ldots,m_{\rm q,\Nf})  \,.
\end{equation}
We remind here that both the necessity of the nonzero critical bare mass
$m_{\rm crit}$ and the extra renormalization constant $r_m$ refer to our
lattice regulator which breaks chiral symmetry.

To implement Symanzik O($a$) improvement the above relations have to be
augmented by additional terms that are proportional to an explicit factor
$a$~\cite{Luscher:1996sc}. In the case of the coupling this is of a relatively
simple form. All we have to do is to replace $g_0^2$ in (\ref{grm0}) and
(\ref{mrm0}) by
\begin{equation}
 \gtilde^2 = g_0^2 \left( 1 + a\bg(g_0^2) \,\tr[\Mq]/\Nf \right) \,,
\label{g0tilde}
\end{equation}
and to add to our lattice action the `clover' contribution, first introduced by
Sheikholeslami and Wohlert (SW)~\cite{Sheikholeslami:1985ij},
\begin{equation}
 S_{\sss\rm SW} = a^5 \csw(g_0^2) \sum_x \bar{\psi}(x) \frac{i}{4} \sigma_{\mu\nu} \hat{F}_{\mu\nu}(x) \psi(x) \,,
 \label{Sym_clover_term}
\end{equation}
with $\hat{F}_{\mu\nu}(x)$ being a lattice discretization of the
(anti-Hermitian) field strength tensor at site $x$.  More details on the
lattice action follow in the next section.  This extra term corresponds to the
contribution proportional to $\sigma_0$ in (\ref{Sym_action}), while the $\bg$
modification in~\eqref{g0tilde} represents the $\sigma_5$ part on the lattice.

Sandwiching the matrix $M+a\{\sigma_1 M^2+\sigma_2 \tr[M]M +\sigma_3
\tr[M^2]+\sigma_4 (\tr[M])^2\}$ between $\bar{\psi}$ and $\psi$ summarizes the
remaining terms in (\ref{Sym_action}).  Following the notation
of~\cite{Bhattacharya:2005rb} the corresponding terms are included in the
lattice theory by using an improved renormalized mass of flavor $i$
\begin{align}
\mR[,i] &= Z_m(\gtilde^2, a\mu) \bigg[ \mqi + \left( r_m(\gtilde^2) - 1 \right) \frac{\tr[\Mq]}{\Nf} + a \bigg\{
        b_m(g_0^2) \mqi^2 + \bar{b}_m(g_0^2) \tr[\Mq]\, \mqi \nonumber\\
  &\qquad\qquad\qquad\qquad\qquad + \left( r_m(g_0^2) d_m(g_0^2) - b_m(g_0^2) \right) \frac{\tr[\Mq^2]}{\Nf} \notag \\
  &\qquad\qquad\qquad\qquad\qquad + \left( r_m(g_0^2) \bar{d}_m(g_0^2) - \bar{b}_m(g_0^2) \right)
      \frac{\left(\tr[\Mq]\right)^2}{\Nf} \bigg\} \bigg]  \label{mri0}
\end{align}
on the lattice.  If now physical observables are parameterized by $\gR$ and
$\mR[,i]$ then we expect them to converge to their continuum limits at a rate
proportional to $a^2$, provided that the prefactors of all improvement terms
have the correct values to cancel all O($a$) terms along with the divergences.

In practice, for $\Nf \le 3$, the mass dependent improvement terms have been
found to lead to small effects. Thus they could be set to their low order
perturbative values without spoiling the targeted precision.  Only the $\csw$
term (beside operator improvements to be discussed later) was found to be
essential.  For this reason the ALPHA collaboration and others have engaged in
nonperturbative computations of $\csw(g_0^2)$ for various numbers of light
flavors $0\le \Nf \le4$. They are found in
refs.~\cite{Luscher:1996ug,Jansen:1998mx,Yamada:2004ja,Aoki:2005et,Cundy:2009yy,Tekin:2009kq,Bulava:2013cta}.

If we want to treat the physical charm quark in a mass independent scheme as
discussed above, then there appear improvement terms multiplied by the charm
mass $a\mqcharm$. Also $\tr[\Mq]$ will now be dominated by charm. A rough
estimate puts these terms to around 0.5 and about an order of magnitude larger
than the analogous strange contributions for lattice spacings on the order of
$0.1\,\fm$. Thus the respective $b_{*}$ and other coefficients must be known
much more precisely than for the light flavors. This is even true for the
virtual contributions in the absence of valence charm quarks. Even at the level
$a^0$ the term $r_m-1$ is enhanced and very high precision for $r_m$ may be
required. A simultaneous nonperturbative determination of all these improvement
coefficients seems impractical and we hence rather give up the technical
convenience of mass independent schemes and switch to massive ones.

\subsection{Mass dependent case}

In massive renormalization schemes the $Z$ factors are allowed to depend on the
mass matrix $M$.  We take
\begin{equation}
 \gR^2 = \tilde{Z}_g(g_0^2,a\mu,aM) g_0^2 \,, \label{grm}
\end{equation}
and
\begin{equation}
 \mR[,i] = \tilde{Z}_m^i(g_0^2,a\mu,aM)\left[m_i-\tilde{m}_{\rm crit}(g_0^2,a\tr[\Mq])\right] \,. \label{mrm}
\end{equation}
Clearly all the correction terms proportional to $a\mqi$ in~\eqref{g0tilde}
and~\eqref{mri0} can now be absorbed into the multiplicative $\tilde{Z}$
factors. The term proportional to $(r_m-1)$ in~\eqref{mri0} can be reproduced
by the $a\tr[\Mq]$ dependent critical mass $\mcrtil$. It remains to keep the
clover term~\eqref{Sym_clover_term} with a mass dependent coefficient function
$\cswtil(g_0^2,aM)$. Note that the mass dependent O($a$) modification of this
improvement term formally only contributes at O($a^2$) and in this sense could
be dropped.  But in passing to mass dependent renormalization we want to avoid
these O($a^2$) effects which are enhanced by the charm mass and keep the mass
dependence of $\cswtil$, too.

Beside the definition of renormalized parameters and the SW term the scheme has
to include improved operators, and we focus on $A_{\mu}^{ij}$ and $P^{ij}$ as
before. The definitions~\eqref{Amucurrent} and~\eqref{Pcurrent} are now
reinterpreted in terms of lattice fields all at the same site.  By similar
considerations as for the action we set
\begin{equation}
  (A_{\sss\rm R})^{ij}_{\mu}  = \ZAtil[ij](g_0^2,aM)\left[ A^{ij}_{\mu} + a\cAtil(g_0^2,aM) \partial_{\mu}P^{ij} \right]
\end{equation}
and
\begin{equation}
(P_{\sss\rm R})^{ij}=\ZPtil[ij](g_0^2,a\mu,aM) P^{ij}
\end{equation}
for the renormalized and improved currents in our massive scheme.%
\footnote{We note that mass independent renormalization schemes have
          turned out to be very practical for QCD with $\nf\leq3$. While the
          large charm mass motivates to renormalize at finite $m_\charm$, it is
          presumably of interest to keep $m_\up=m_\down=m_\strange=0$, defining
          all $Z$-factors at such a point where less parameters have to be
          tuned. In practice this means that renormalization factors and
          improvement coefficients are defined at the point
          $\Mq=(0,0,0,\mqi[\charm])$ and are thus functions of $a\mqi[\charm]$.
          At this  point they can be determined exactly as in an $\Nf=3$
          theory, as long as the fields do not involve the charm quark. The
          small $\rmO(a\mqi[l])$ change in $\csw$ amounts to a negligible
          overall $\rmO(a^2)$ term. For fields containing charm, special
          renormalization conditions have to be imposed.
}%
For the inclusion of an $M$ dependence in $\cAtil$ we refer to the earlier
discussion for $\cswtil$.

\subsection{A 3+1 flavor scheme}

So far we have assumed no degeneracies in our arbitrary quark mass matrix.  For
our later numerical work we shall find it advantageous to reduce the number of
relevant mass parameters by approximating nature by a simplified pattern. We
shall be guided in principle by keeping the charm mass and the value of
$\tr[\Mq]$ close to their physical values. At the same time we deal with only
two independent masses by setting
\begin{equation}
 \Mq = \diag(\mqi[l], \mqi[l], \mqi[l], \mqi[\charm] ) \,.
 \label{e:Mqsu3}
\end{equation}
Beside the (subtracted) charm mass $\mqi[\charm]$ we have introduced a light
quark mass $\mqi[l]$ which is imagined to equal approximately a third of the
strange mass.  We expect this scheme to be at the same time numerically
manageable and close enough to the eventual target physics to avoid large mass
dependent artefacts and renormalization effects.

In the following we shall be concerned with this kind of massive
renormalization scheme and will omit again the tildes over the improvement
coefficients $\cA, \csw$. Moreover we shall in particular embark on a
nonperturbative calculation of $\csw(g_0^2,a\Mq)$ with $\Mq$ chosen as in
\eq{e:Mqsu3}.

\section{Scaling at constant finite volume physics}\label{sec:LCPscaling}

To prepare for a nonperturbative determination of $\csw$ we now envisage the
construction of a set of lattice configurations with bare parameters that
correspond to a line of constant physics (LCP) with fixed renormalized
parameters in a finite volume scheme and a well defined discretization.  This
is done for a number of lattice spacings $a$ and $\csw$ values in the relevant
range. The quark masses for $\Nf=3+1$ are tuned to the vicinity of the values
described at the end of the last section.  In the next section we consider
scaling violations in the PCAC relation (\ref{PCAC0}) to single out `optimal'
$\csw$ values for each of our lattice spacings which can then be represented
and interpolated by a smooth fit formula.

We implement Schr\"odinger functional (SF) boundary
conditions~\cite{Luscher:1992an,Sint:1993un} to set up our finite volume
scheme.  The 3+1 flavors of Wilson quarks are coupled to gluons weighted with
the tree level improved L\"uscher--Weisz gauge action~\cite{Luscher:1984xn}.
Instead of compiling all details of these definitions we refer here
to~\cite{Bulava:2013cta} where almost the same setup has been used.  The only
small difference is that for the fermionic boundary improvement coefficient
$\cttil$, that is called $c_\mathrm{\sss F}$ in~\cite{Bulava:2013cta}, we use
the tree level approximation $\cttil=1$.  We choose equal temporal and spatial
extents $T=L$.  The boundary gauge fields $C_k, C_k'$ will first be set to zero
in this section and to the nontrivial values given
in~\cite{Bulava:2013cta,Luscher:1996ug} in the following section. Our fermion
fields are periodic in the space direction, i.e., the freedom to insert a
periodicity angle $\theta$ into the SF is not used here.  The bare quark masses
will be specified in terms of the well known hopping parameters
\begin{align}
 \kappa_l      &= \frac{1}{2(am_{l}+4)}      \,, & 
 \kappa_\charm &= \frac{1}{2(am_{\charm}+4)} \,.
\end{align}

\subsection{Finite size definition of the LCP}

We start with the definition of a renormalized coupling that we denote by
$\Phi_1$.  In recent years it has turned out that coupling constants derived
from the gradient flow~\cite{Narayanan:2006rf,Luscher:2010iy} have an excellent
signal quality for finite tori~\cite{Fodor:2012td,Ramos:2014kla} and for SF
boundary conditions~\cite{Fritzsch:2013je} at the range of system sizes that we
shall want to simulate. We employ the coupling given by the expectation value
of the action density
\begin{equation}
        \Phi_1 =  \frac{t^2}{\mathcal{N}} \sum_{\mu \nu}\left\langle -\tfrac{1}{2}\tr[G_{\mu \nu}^2(x,t)]  \right\rangle\Big|_{\scalebox{0.7}{$x_0=T/2,t=c^2 L^2/8$}} \,.
\end{equation}
In this formula $G_{\mu \nu}$ is the (anti-Hermitian traceless) clover field
strength of the `flowed' field and its flow time argument $t$ is tied to the
system size by
\begin{equation}
 c = 0.3 \,.
\end{equation}
The lattice flow equation is based on the simple Wilson plaquette action, i.e.,
we work with the Wilson flow.  The normalization factor ${\cal N}$, which
implies $\Phi_1=g_0^2+{\rm O}(g_0^4)$ in perturbation theory, has been derived
in~\cite{Fritzsch:2013je}.

Apart from the coupling that runs with the system size $T=L$ we need two more
dimensionless renormalized quantities to fix the light and charm masses.  To
that end we introduce the SF correlations
\begin{align}
        \fA^{ij}{ (x_0)} &= -\frac{1}{2} \big\langle A_{0}^{ij}{ (x_0)} {\cal O}^{ji} \big\rangle \,, &
        \fP^{ij}{ (x_0)} &= -\frac{1}{2} \big\langle P^{ij}{ (x_0)} {\cal O}^{ji}     \big\rangle \,,
 \label{SFcorr}
\end{align}
with the lattice currents (\ref{Amucurrent}). The SF boundary
operator~\cite{Luscher:1996sc}
\begin{equation}
        {\cal O}^{ij} = a^6 \sum_{\vec{x}, \vec{y}} \bar{\zeta}^{ i}(\vec{x}) \gamma_5 \zeta^{ j}(\vec{y})
\end{equation}
is built from boundary fields at $x_0=0$ and yields a nonzero correlation.  We
next form the improved axial correlation
\begin{equation}
 \fAI^{ij} = \fA^{ij} + a\cA \tilde{\partial}_0  \fP^{ij}
\end{equation}
involving the symmetric lattice derivative $\tilde{\partial}_0$.  An effective
meson mass times system size is now coded into
\begin{equation}
 \Phi_2 = -\tilde{\partial}_0 \log\!\big( \fAI^{12}(x_0) \big) \big|_{x_0=T/2} \times T \,.
\end{equation}
For very large $T$ we would isolate the mass of the pseudoscalar non-singlet
meson of two different light quark species.  At finite $T$ we instead have a
combination of this mass and higher levels in the same channel with weights
given by universal amplitude ratios \cite{Guagnelli:1999zf}.  In a third
quantity we excite states with the quantum number of a singly charmed meson
(fourth flavor) and take
\begin{equation}
        \Phi_3 = -\tilde{\partial}_0 \log\!\big( \fAI^{14}(x_0) \big) \big|_{\sss x_0=T/2} \times T - \tfrac{1}{2}  \Phi_2 \,.
\end{equation}
The reason for the subtraction of $\Phi_2$ is as follows. We will have to solve
the complicated tuning task to find values of $g_0^2,\kappa_l,\kappa_\charm$
that produce certain soon to be prescribed values of $\Phi_{1,2,3}$. This task
is facilitated if $\Phi_3$ is sensitive to $\kappa_\charm$ and shows only weak
dependence on $\kappa_l$. The subtraction cancels the latter dependence in the
most naive constituent quark model. In $\Nf=2$ simulations it has been verified
that this property holds in an approximate sense also beyond such a naive
scenario. 

Another remark is in order to why we use a component of the axial vector rather
than the pseudoscalar current to excite the desired quantum numbers. At large
$T$ the same meson mass would be isolated with the admixtures at finite $T$
being different in the two cases.  The PCAC relation implies that our $\Phi_2$
vanishes in the chiral limit, while this would not be the case for the
analogous quantity derived with the $P^{12}$ correlator.  Hence our choice is
expected to optimize the tuning sensitivity for the light mass $\kappa_l$.
Finally also the inclusion of the improvement term (with perturbative 1-loop
$\cA$~\cite{Aoki:1998qd}) in $\fAI$ is not a necessity here and, again at
$\Nf=2$, it has been seen to make little effect.  Nevertheless we keep it as
part of our choice.

\subsection{Values of $\Phi_{1,2,3}$}

Although not practical, a `Gedanken simulation' may be useful to appreciate the
well-defined status of our finite volume quantities. We may imagine to simulate
QCD on a lattice with negligible cutoff and finite size effects where the bare
lattice parameters are tuned such that pion, kaon, D-meson masses and a scale
such as the pion decay constant equal their values compiled by the particle
data group. From the resulting physical values of the renormalization group
invariant (RGI) quark masses, $M_\up,\,M_\down,\,M_\strange, \,M_\charm$ we
could then easily switch to a flavor SU(3) symmetric point
$M_l=(M_\up+M_\down+M_\strange)/3$ for the three light species keeping
$M_\charm$.  Then $a$ would be known in physical units and we may lower
$T/a=L/a$ such that we find $L\approx 0.8\,\fm$, for example, still having
$L/a\gg 1$.  On the resulting lattice we read off universal continuum values
for $\Phi_{1,2,3}$.

The above strategy is, of course, impractical. Instead we have to somehow
deduce the $\Phi_i$ with sufficient precision, in order to then on a series of
LCP lattices tune $\csw$ for O($a$) improvement and to later use this
information to reach acceptably small cutoff effects in reasonably large volume
simulations. The choice of the system size $L$ together with the feasible $L/a$
determine for which $g_0^2$ we obtain nonperturbative information on $\csw$. It
should end up in a range useful for the final simulations.

Our strategy to determine $\Phi_i$ is to use here the approximation $\Nf=2$
with quenched strange and charmed quarks and configurations from earlier
simulations%
\footnote{Their availability was another argument to choose an extent around
          $L\approx 0.8\,\fm$ for the LCP.
}%
of the ALPHA collaboration~\cite{Fritzsch:2012wq,Blossier:2012qu} together with
additional measurements and some newly created configurations as well.  For
details on these well-defined but tedious estimations we refer
to~\cite{Thesis:Stollenwerk}.  We here just report the outcome in the form of
values
\begin{equation}
 \Phi_1^\ast = 7.31,\quad \Phi_2^\ast = 0.59,\quad \Phi_3^\ast = 5.96.
 \label{Phitarget}
\end{equation}
These numbers emerge from the $\Nf=2$ computation with errors at the percent
level%
\footnote{Not including partial quenching errors, of course.
}%
but are now taken to {\em define} our LCP which in the continuum limit yields a
finite slab of continuum QCD with SF boundary conditions.

An important question is now to which precision we have to tune our $\Nf=3+1$
lattices to the above target values $\Phi_i^\ast$ to then determine
$\csw(g_0^2)$ along the LCP. By a chain of heuristic considerations given in
more detail in~\cite{Thesis:Stollenwerk} we arrive at the requirement
\begin{align}
 \frac{\Delta\Phi_1}{\Phi_1^\ast} &\lesssim 4\%  \,, &
 \frac{\Delta\Phi_2}{\Phi_2^\ast} &\lesssim 10\% \,, &
 \frac{\Delta\Phi_3}{\Phi_3^\ast} &\lesssim 4\%  \,.
\end{align}
This corresponds roughly to 5\% precision in the scale $T$ and the charm mass
and 11\% in the light mass. As discussed earlier the closeness to the physical
charm mass is of particular relevance to avoid large cutoff effects
proportional to the deviation $\Delta\mqi[\charm]$. The precise values of the
light quark masses, which are dominantly controlled by $\Phi_2$, on the other
hand appear to be less  critical.

\subsection{Lattice realization of the LCP}

We now report on a set of simulation results with lattices tuned to
$\Phi_i^\ast$ within the required precision. They were found from a first set
of exploratory simulations in the right range followed by multidimensional
interpolations.  They serve to predict the `right' values of
$g_0,\kappa_l,\kappa_\charm$ which are then confirmed or fed into refined
interpolations. The final results from this somewhat demanding procedure
(see~\cite{Thesis:Stollenwerk} for more details) are collected in
table~\ref{tab:914}.  Note that a range of $\csw$ is covered for each value
$T^\ast\!/a$.

After some initial tests, all simulations have been performed in a
(2+1+1)-flavor setup, i.e., a degenerate doublet of light quarks is simulated
by HMC~\cite{Duane:1987de} and the two remaining quark species are incorporated
by the RHMC algorithm~\cite{Kennedy:1998cu,Clark:2006fx}.  In general, the
simulations are conducted along the lines of Ref.~\cite{DallaBrida:2016kgh},
appendix~A, and more details and a cost figure will be
provided~\cite{Nf3tuning}.  Due to the enhanced spectral gap in the Dirac
operator with SF boundary conditions and non-vanishing physical quark masses,
the performed HMC simulations are very stable and unproblematic.

%
%
\begin{table}[!t]
   \small
   \centering
   \begin{tabular}{LLLLLLL}\toprule
     \csw & g_0^2  &\kappa_l & \kappa_\charm & \Phi_1  & \Phi_2  &  \Phi_3  \\\midrule
      \multicolumn{7}{c}{$T^\ast\!/a=8$}                               \\\cmidrule(lr){1-7}
      1.9 & 1.7848 & 0.13741 & 0.12033  & 7.17(4) & 0.59(3) &  5.95(1) \\
      2.0 & 1.8067 & 0.13645 & 0.11960  & 7.17(3) & 0.56(2) &  5.99(1) \\
      2.1 & 1.8340 & 0.13560 & 0.11900  & 7.36(5) & 0.61(3) &  6.02(1) \\
      2.2 & 1.8534 & 0.13450 & 0.11820  & 7.17(5) & 0.60(3) &  6.03(1) \\
      2.3 & 1.8825 & 0.13369 & 0.11770  & 7.36(3) & 0.62(2) &  5.97(1) \\
      2.4 & 1.9069 & 0.13278 & 0.11750  & 7.31(4) & 0.57(3) &  5.93(1) \\\cmidrule(lr){1-7}
      \multicolumn{7}{c}{$T^\ast\!/a=12$}                              \\\cmidrule(lr){1-7}
      1.8 & 1.7334 & 0.13742 & 0.12776  & 7.13(5) & 0.58(3) &  5.93(2) \\  
      1.9 & 1.7600 & 0.13656 & 0.12714  & 7.26(4) & 0.60(3) &  5.92(2) \\
      2.0 & 1.7876 & 0.13573 & 0.12653  & 7.35(5) & 0.62(3) &  5.99(3) \\
      2.1 & 1.8152 & 0.13486 & 0.12591  & 7.29(6) & 0.60(4) &  5.93(3) \\
      2.2 & 1.8423 & 0.13401 & 0.12529  & 7.42(7) & 0.60(4) &  5.91(3) \\\cmidrule(lr){1-7}
      \multicolumn{7}{c}{$T^\ast\!/a=16$}                              \\\cmidrule(lr){1-7}
      1.5 & 1.6266 & 0.13918 & 0.13224  & 7.38(9) & 0.56(2) &  5.93(3) \\
      1.6 & 1.6482 & 0.13825 & 0.13139  & 7.42(10)& 0.61(2) &  6.06(7) \\
      1.7 & 1.6741 & 0.13741 & 0.13087  & 7.29(8) & 0.59(2) &  5.84(5) \\
      1.8 & 1.7036 & 0.13661 & 0.13018  & 7.32(7) & 0.57(2) &  5.90(5) \\
      1.9 & 1.7331 & 0.13582 & 0.12949  & 7.44(9) & 0.59(2) &  5.94(3) \\
      2.0 & 1.7626 & 0.13503 & 0.12880  & 7.46(7) & 0.59(2) &  6.03(4) \\\cmidrule(lr){1-7}
      \multicolumn{7}{c}{$T^\ast\!/a=20$}                              \\\cmidrule(lr){1-7}
      1.5 & 1.5868 & 0.13822 & 0.13287  & 7.20(7) & 0.61(2) &  5.91(8) \\
      1.6 & 1.6153 & 0.13749 & 0.13240  & 7.30(10)& 0.57(2) &  5.91(5) \\
      1.7 & 1.6434 & 0.13669 & 0.13178  & 7.29(7) & 0.60(2) &  5.85(5) \\
      1.8 & 1.6714 & 0.13588 & 0.13126  & 7.49(5) & 0.63(2) &  5.81(6) \\
      1.9 & 1.6993 & 0.13504 & 0.13046  & 7.36(9) & 0.62(2) &  5.81(8) \\\cmidrule(lr){1-7}
      \multicolumn{7}{c}{$T^\ast\!/a=24$}                              \\\cmidrule(lr){1-7}
      1.5 & 1.5527 & 0.13751 & 0.13325  & 7.19(6) & 0.62(2) &  6.02(8) \\
      1.6 & 1.5783 & 0.13672 & 0.13253  & 7.23(11)& 0.59(1) &  6.02(6) \\
      1.7 & 1.6065 & 0.13595 & 0.13202  & 7.24(9) & 0.62(2) &  6.00(5) \\
      1.8 & 1.6344 & 0.13515 & 0.13132  & 7.11(7) & 0.56(1) &  5.87(6) \\
     \bottomrule
   \end{tabular}
   \caption{Results for tuning to the line of constant physics,
            eq.~\eqref{Phitarget}, at various values of $\csw$.
           }
   \label{tab:914}
\end{table}

\section{Determination of $\csw$}\label{sec:csw}

%
%
\begin{table}[!t]
   \small
   \centering
   \begin{tabular}{LLLLLLL}\toprule
    \csw & g_0^2  &\kappa_l &\kappa_\charm & T^*m^{ud} & T^*M^{ud} & T^*\Delta M^{ud} \\\midrule
      \multicolumn{7}{c}{$T^*\!/a= 8$}                                     \\\cmidrule(lr){1-7}
     1.9 & 1.7848 & 0.13741 & 0.12033 & 0.504(7)  & 0.684(25) & +0.047(7)  \\  
     2.0 & 1.8067 & 0.13645 & 0.11960 & 0.491(8)  & 0.621(23) & +0.028(7)  \\
     2.1 & 1.8340 & 0.13560 & 0.11900 & 0.509(8)  & 0.707(32) & +0.021(8)  \\
     2.2 & 1.8534 & 0.13450 & 0.11820 & 0.516(8)  & 0.643(28) & -0.014(8)  \\
     2.3 & 1.8825 & 0.13369 & 0.11770 & 0.521(9)  & 0.724(30) & -0.013(8)  \\
     2.4 & 1.9069 & 0.13278 & 0.11750 & 0.521(10) & 0.723(31) & -0.016(9)  \\\cmidrule(lr){1-7}
      \multicolumn{7}{c}{$T^*\!/a=12$}                                     \\\cmidrule(lr){1-7}
     1.8 & 1.7334 & 0.13742 & 0.12776 & 0.247(4)  & 0.311(11) & +0.032(4)  \\
     1.9 & 1.7600 & 0.13656 & 0.12714 & 0.242(4)  & 0.306(9)  & +0.012(7)  \\
     2.0 & 1.7876 & 0.13573 & 0.12653 & 0.237(4)  & 0.300(11) & +0.020(5)  \\
     2.1 & 1.8152 & 0.13486 & 0.12591 & 0.243(5)  & 0.300(10) & -0.004(7)  \\
     2.2 & 1.8423 & 0.13401 & 0.12529 & 0.223(4)  & 0.267(11) & -0.011(6)  \\\cmidrule(lr){1-7}
      \multicolumn{7}{c}{$T^*\!/a=16$}                                     \\\cmidrule(lr){1-7}
     1.5 & 1.6266 & 0.13918 & 0.13224 & 0.160(3)  & 0.192(6)  & +0.026(3)  \\
     1.6 & 1.6482 & 0.13825 & 0.13139 & 0.159(2)  & 0.164(5)  & +0.018(4)  \\
     1.7 & 1.6741 & 0.13741 & 0.13087 & 0.156(2)  & 0.139(6)  & +0.014(3)  \\
     1.8 & 1.7036 & 0.13661 & 0.13018 & 0.167(2)  & 0.144(6)  & +0.004(3)  \\
     1.9 & 1.7331 & 0.13582 & 0.12949 & 0.165(3)  & 0.136(6)  & -0.008(3)  \\
     2.0 & 1.7626 & 0.13503 & 0.12880 & 0.154(2)  & 0.113(6)  & -0.015(4)  \\\cmidrule(lr){1-7}
      \multicolumn{7}{c}{$T^*\!/a=20$}                                     \\\cmidrule(lr){1-7}
     1.5 & 1.5868 & 0.13822 & 0.13287 & 0.150(2)  & 0.153(5)  & +0.018(4)  \\
     1.6 & 1.6153 & 0.13749 & 0.13240 & 0.131(2)  & 0.135(5)  & +0.006(3)  \\
     1.7 & 1.6434 & 0.13669 & 0.13178 & 0.139(2)  & 0.117(5)  & -0.000(3)  \\
     1.8 & 1.6714 & 0.13588 & 0.13126 & 0.132(3)  & 0.097(5)  & -0.003(3)  \\
     1.9 & 1.6993 & 0.13504 & 0.13046 & 0.140(2)  & 0.093(5)  & -0.007(5)  \\\cmidrule(lr){1-7}
      \multicolumn{7}{c}{$T^*\!/a=24$}                                     \\\cmidrule(lr){1-7}
     1.5 & 1.5527 & 0.13751 & 0.13325 & 0.129(2)  & 0.135(4)  & +0.007(4)  \\
     1.6 & 1.5783 & 0.13672 & 0.13253 & 0.120(1)  & 0.109(3)  & +0.011(2)  \\
     1.7 & 1.6065 & 0.13595 & 0.13202 & 0.119(1)  & 0.092(3)  & -0.004(2)  \\
     1.8 & 1.6344 & 0.13515 & 0.13132 & 0.113(2)  & 0.076(4)  & -0.007(2)  \\
     \bottomrule
   \end{tabular}
   \caption{Results of the individual improvement condition runs along the LCP for 
            various input values of $\csw$. Beside $g_0$, $\kappa_l$ and
            $\kappa_\charm$, we quote the corresponding current quark masses
            $m^{ud}(x_0)$ at $x_0=T^*/2$, the improvement condition mass
            $M^{ud}(x_0,y_0)$ for $x_0=\tfrac{1}{4}T^*$ and
            $y_0=\tfrac{3}{4}T^*$, and the mass difference $M^{ud}$ from
            eq.~\eqref{eq:Mcondition}. Errors from the relation $g_0^2
            \leftrightarrow T^*/a$ are neglected.
           }
   \label{tab:915}
\end{table}

For each of the groups of LCP configurations at a given $T^\ast\!/a$ we now
want to find the combinations of $g_0^2$ and $\csw$ that lead to the
cancellation of a certain lattice artefact. Equivalently we can say that we
impose an improvement condition where we match a result to the continuum. As we
indicated before the PCAC operator relation is a source of such conditions
which has in fact been traditionally used~\cite{Luscher:1996ug,Jansen:1998mx}
to determine $\csw$.  To achieve a good sensitivity to $\csw$ a chromoelectric
background field is enforced in the SF by the nonvanishing boundary fields
$C_k, C'_k$ of~\cite{Luscher:1996ug}.

The bare improved current (PCAC) quark mass is written in terms of the improved
SF correlations~\eqref{SFcorr}%
\begin{equation}
  m^{ij}(x_0)= r^{ij}(x_0) + a\cA s^{ij}(x_0)
\end{equation}
with
\begin{align}
 r^{ij}(x_0) &= \frac{        \tilde{\partial}_0 \fA^{ij}(x_0)}{2\fP^{ij}(x_0)}  \,, &
 s^{ij}(x_0) &= \frac{\partial_0^\ast \partial_0 \fP^{ij}(x_0)}{2\fP^{ij}(x_0)}  \,,
\end{align}
where $\partial_0$, $\partial_0^\ast$ are the forward and backward lattice
difference operators.  The $x_0$ dependence is, of course a lattice artefact,
absent for a true operator relation in the continuum. A second possibility is
given by inserting SF operators ${\cal O}'^{ij}$ in (\ref{SFcorr}) which are
analogous to ${\cal O}^{ij}$ but defined~\cite{Luscher:1996sc} on the opposite
boundary at $x_0=T$.  We add a prime to the corresponding correlations and
finally to ${m'}^{ij}(x_0)$.  We are now ready to state our \emph{improvement
conditions}
\begin{align}
    m^{ij}(x_0) &= {m'}^{ij}(x_0) &
    \text{ at } x_0 &= \tfrac{1}{4} T \text{ and } x_0 = \tfrac{3}{4} T \,.
\end{align}
As we {\em equate} different versions of $m^{ij}$ we never need renormalization
$Z$ factors.  The two Euclidean time locations in $x_0$ are chosen to stay away
from the boundaries for smaller higher order cutoff effects but also from each
other to have two reasonably independent conditions.  Two conditions are
imposed to determine $\csw$ and simultaneously $\cA$.  More precisely we first
determine
\begin{align} 
   \cA = \frac{{r'}^{ij}(y_0)-r^{ij}(y_0)}{{s'}^{ij}(y_0)-s^{ij}(y_0)}  \quad \mbox{ at } y_0 = \tfrac{3}{4} T \,,
\end{align}
and then insert this value into
\begin{align}
    M^{ij} &= r^{ij}(x_0)      + \cA s^{ij}(x_0)     \,, &
    {M'}^{ij} &= {r'}^{ij}(x_0)+ \cA{s'}^{ij}(x_0)   \,, & 
    \mbox{ at } x_0 &= \tfrac{1}{4} T \,.
\end{align}
Our condition to determine $\csw$ finally reads
\begin{equation}\label{eq:Mcondition}
 0 = \Delta M^{ij} = M^{ij}-{M'}^{ij}.
\end{equation}
A closer inspection discloses that the above construction is actually symmetric
under interchanging the insertion times $x_0$ and $y_0$.  So far we left the
quark species $i\not=j$ general. As we shall at first primarily be interested
in light quarks, we shall in the following take $ij=12=ud$.

\begin{figure}[t]
   \small
   \centering
   \includegraphics[height=0.235\textheight]{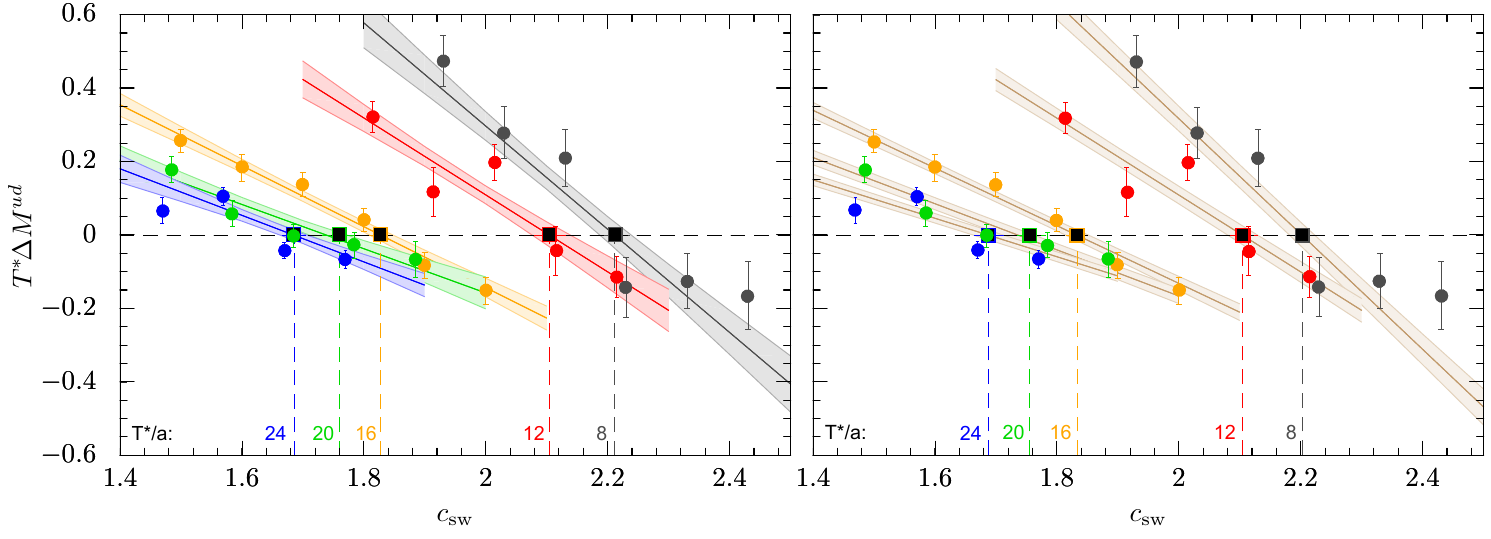}%
   \caption{Interpolations of data from table~\ref{tab:915} to $T^\ast\!\Delta
            M^{ud}=0$ individually at each $T^\ast\!/a$ (left), or using the
            global ansatz of eq.~\eqref{DeltaMfit} (right). Results for $c_{\rm
            sw,I}$ are summarized in table~\ref{tab:916}.
           }
   \label{fig:913}
\end{figure}

%
%
\begin{table}[!t]
   \small
   \centering
   \begin{tabular}{CLLLLLLLL}\toprule
     & \multicolumn{4}{c}{individual fit results}   & \multicolumn{4}{c}{global fit results}  \\\cmidrule(lr){2-5}\cmidrule(lr){6-9}
      T^\ast\!/a 
          & c_{\rm sw,I}
                      & g_0^2  &\kappa_l & \kappa_\charm & c_{\rm sw,I} 
                                                                & g_0^2  &\kappa_l & \kappa_\charm \\\cmidrule(lr){2-5}\cmidrule(lr){6-9}
        8 & 2.211(25) & 1.8597 & 0.13450 & 0.11836  & 2.202(21) & 1.8575 & 0.13459 & 0.11841  \\
       12 & 2.105(30) & 1.8163 & 0.13482 & 0.12588  & 2.105(24) & 1.8163 & 0.13482 & 0.12588  \\
       16 & 1.828(19) & 1.7129 & 0.13640 & 0.12997  & 1.833(19) & 1.7144 & 0.13636 & 0.12993  \\
       20 & 1.760(34) & 1.6600 & 0.13619 & 0.13140  & 1.755(28) & 1.6586 & 0.13623 & 0.13143  \\
       24 & 1.686(20) & 1.6028 & 0.13605 & 0.13205  & 1.688(24) & 1.6035 & 0.13603 & 0.13204  \\
     \bottomrule
   \end{tabular}
   \caption{Interpolation results of $\csw$ at $T^*M^{ud}=0$ for individual
            (local) and a global fit ansatz.
           }
   \label{tab:916}
\end{table}

\begin{figure}[t]
   \small
   \centering
   \includegraphics[width=\textwidth]{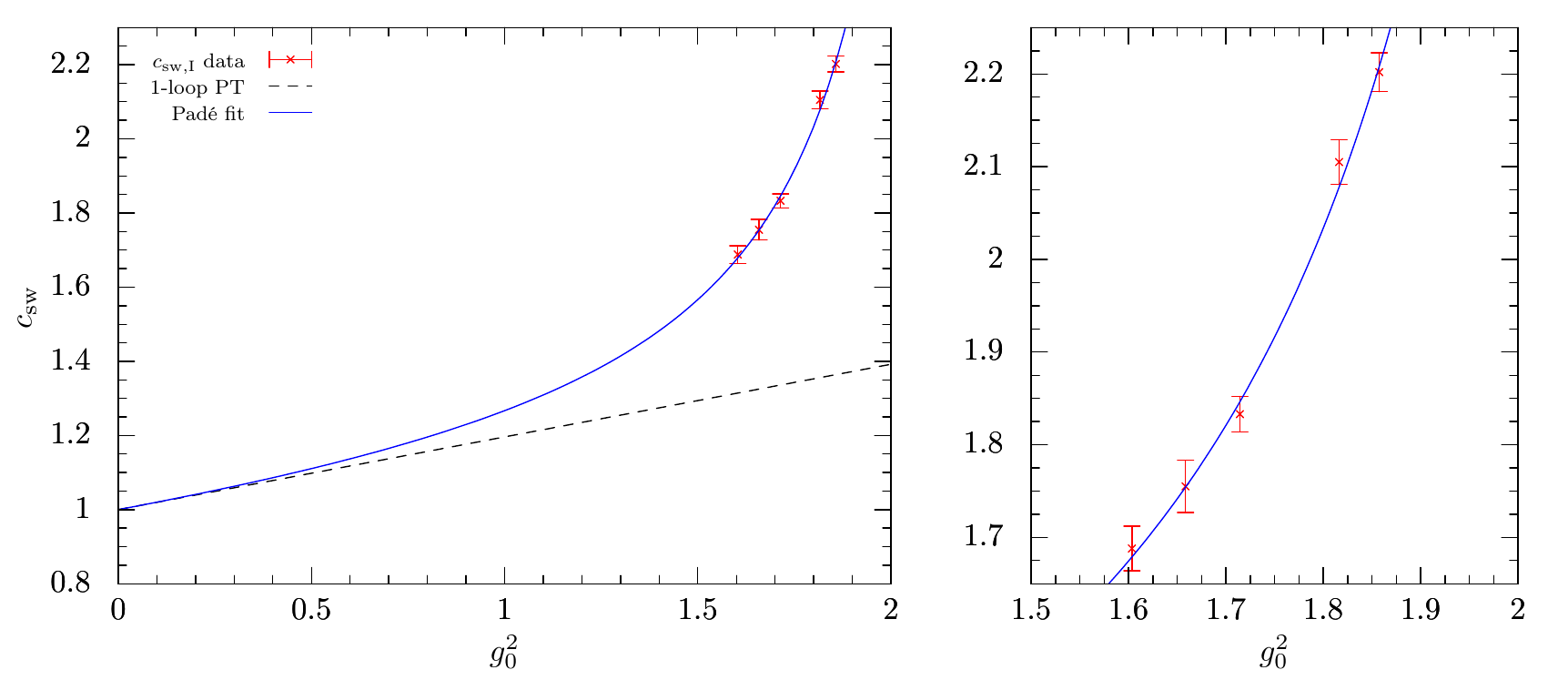}%
   \caption{$c_{\rm sw,I}$ from global fit result as a function of the bare
            gauge coupling $g_0^2$, cf. eq.~\eqref{cswfit}, along our line of
            constant physics.
           }
   \label{fig:915}
\end{figure}

We have determined the bare parameters for LCP lattices with $T=L$. To enhance
the sensitivity to our improvement conditions we shall in the following however
switch to the smaller spatial extent $L=T/2$ and asymmetric lattices.  The
$T/a$ values are unchanged to accommodate the various insertion time slices.
The results for $M^{ud}$ and $\Delta M^{ud}$ on these lattices are listed in
table~\ref{tab:915}.  A glance at the last column shows that we have managed
to bracket a zero of $\Delta M^{ud}T^\ast$ for each value $T^\ast\!/a$. To
precisely locate the zeroes we use the following fit ansatz
\begin{equation}
 T^\ast \Delta M^{ud} = {\rm O}(a) = s \frac{a}{T^\ast}[\csw-c_{\rm sw,I}(g_0^2)] \,.
 \label{DeltaMfit}
\end{equation}
This fit is global in $T^\ast\!/a$ with one slope parameter $s$ and the
improvement values $\csw$ for each of our lattice sizes. It works very well
with $s=-1.26(8)$ as seen in the right panel of figure~\ref{fig:913}.  The
corresponding bare parameters $g_{0}^2, \kappa_{l\rm ,I}, \kappa_{\rm
\charm,I}$ are determined by linearly interpolating the data in
table~\ref{tab:915}. The resulting improvement values are compiled in
table~\ref{tab:916}.  In~\cite{Thesis:Stollenwerk} it is discussed that fits of
the type (\ref{DeltaMfit}) with independent slopes for each $T^\ast\!/a$ lead
to almost the same results.  They are shown in the left panel of
figure~\ref{fig:913}.

We now have a number of values $c_{\rm sw,I}$ for $\csw$ that obey an
improvement condition for five different values $g_0^2$ corresponding to
$T^\ast\!/a=8,12,16,20,24$.  In a last step we represent these data by a
continuous fit function that interpolates between our `sampling' points and
matches the one-loop perturbative result~\cite{Aoki:1998qd}
\begin{equation}
 \csw = 1+0.196 g_0^2 + {\rm O}(g_0^4) \,,
\end{equation}
for $g_0^2 \to 0$.  Our fit is given by the Pad{\'e} form
\begin{align}
 c_{\rm sw,I}(g_0^2) &= \frac{1+A g_0^2+B g_0^4}{1+(A-0.196)g_0^2} \,, & 
 A &= -0.257 \,, &
 B &= -0.050 \,.
 \label{cswfit}
\end{align}
The fit parameters emerge with errors, of course, but at this point we neglect
this and propose the fit formula as a {\em definition} of an O($a$) improved
action.  In figure~\ref{fig:915} we see that the data are represented well. An
initially used $g_0^6$ term in the numerator turned out not to be needed and
receive a coefficient compatible with zero.  In a few tests we finally have
convinced~\cite{Thesis:Stollenwerk} ourselves that the spreads that we have
allowed in tuning the $\Phi_i^\ast$ are subdominant in the errors of the fitted
$\csw$.

\section{Conclusion}

We have emphasized that the large mass $m_\charm$ of charmed quarks leads to
lattice artefacts proportional to $am_\charm \approx 0.5$ for presently
attainable lattice spacings. In the usual setting of a mass independent
renormalization scheme with O($a$) Symanzik improvement $am_\charm$ appears in
numerous improvement terms, which are an order more important than the
analogous light quark terms. For unquenched charm even observables in the light
quark sector are polluted by such cutoff effects.

We find it impractical to tune all corresponding coefficients precisely enough,
but instead propose and design a scheme of renormalization and improvement
conditions formulated at or close to the physical charm mass.  With
Schr\"odinger functional boundary conditions we introduce a renormalized
coupling and effective meson masses that define a finite size scheme. Held
fixed, they define a series of lattices where only the lattice spacing changes.
We use them to compute the coefficient $\csw(g_0^2)$ of the clover improvement
term.  Our result is wrapped up in the formula (\ref{cswfit}) that we would
like to advocate for use in future dynamical charm quark simulations.

\begin{acknowledgement}%
We thank O.~B\"ar and T.~Korzec for input and discussions at an early stage of
this project and P.~Weisz for a critical reading of the manuscript. We are
indebted to our colleagues in the ALPHA collaboration, especially those who
worked on the code basis when this project started: S.~Schaefer, H.~Simma and
A.~Ramos for their modifications of the production code (an extension of
\openQCD\texttt{v1.0}~\cite{openQCD,Luscher:2012av}) and C.~Wittemeier for the
\SFCF~\cite{master:Wittemeier} code to measure SF correlation functions. Last
but not least, the first determination of $\csw$ along a line of constant
physics would not have been possible without the generous computer support at
HLRN (bep00040), NIC at DESY Zeuthen (PAX) and the AG COM at HU.  This work was
supported by the Deutsche Forschungsgemeinschaft in the SFB/TR 09.
\end{acknowledgement}

\small
\bibliographystyle{JHEP_erratum}
\bibliography{mainbib}

\end{document}